# Start-to-end simulation of the injector for a compact THz source

LI Ji(李骥)[1], PEI Yuan-ji(裴元吉)[1], Shang Lei (尚雷)[1*], FENG Guang-yao(冯光耀)[3],
HU Tong-ning(胡桐宁)[2], CHEN Qu-shan(陈曲珊)[2], Li Cheng-long(李成龙)[1]

[1] *National Synchrotron Radiation Laboratory, University of Science and Technology of China, Hefei 230029, China;*
[2] *State Key Laboratory of Advanced Electromagnetic Engineering and Technology, Huazhong University of Science and Technology, Wuhan 430074, China;*
[3] *Deutsches Elektronen-Synchrotron, Hamburg 22607, Germany*

**Abstract:** Terahertz radiation has broad application prospect due to its ability to penetrate deep into many organic materials without damage caused by ionizing radiations. A FEL-based THz source is the best choice to produce high-power radiation. In this paper, a 14 MeV injector is introduced for generating high-quality beam for FEL, which is composed of an EC-ITC RF gun, compensating coils and a travelling-wave structure. Start-to-end simulation has been done with ASTRA code to verify the design and to optimize parameters. Simulation of operating mode at 6 MeV is also executed.

## 1  Introduction

Terahertz wave has very important academic and application value. Compared to other Terahertz sources, FEL is the best way to get maximum power output [1]. In the long term, compact THz sources have broader application prospect. HUST (Huazhong University of Science and Technology) and NSRL (National Synchrotron Radiation Laboratory) / USTC (University of Science and Technology) have jointly proposed a compact THz radiation source based on FEL, which is on construction now. The layout of the whole facility is shown as fig.1 [2].

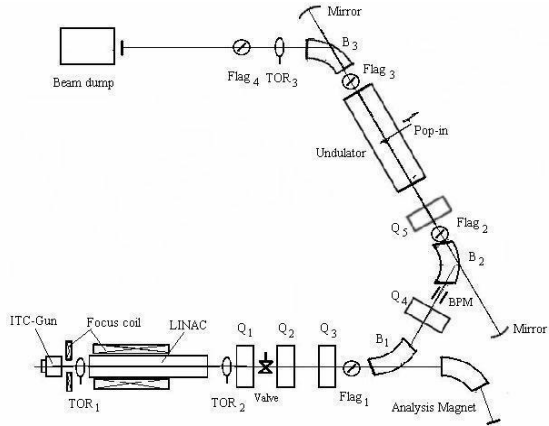

Figure 1: Layout of the THz source facility

THz source requires a high-quality electron source, and photocathode RF gun is the best choice [3]. But the complicated laser drive system and low quantum efficiency of photocathode RF gun make the thermionic RF gun also have broad application prospect. NSRL has developed a thermionic RF gun with two independently tunable cells (ITC) and the external-cathode (EC) structure for years, which is called EC-ITC RF gun. The EC-ITC RF gun can generate beam bunches with superior characteristics for THz source [4].

The beam from EC-ITC RF gun is accelerated to 6~14 MeV in the constant gradient traveling-wave structure, and then transported to the undulator through the transport line. A short magnetic lens and a solenoid were designed to compensate the emittance growth in the injector. The main parameters of the injector are shown in Tab. 1.

Table 1  Expected properties of output beam from the injector

| | |
|---|---|
| Beam energy (effective part ) | 6~14 MeV |
| Beam current (micro pulse) | > 30 A |
| Micro-pulse length (FWHM) | 1~10 ps |
| Energy spread (effective part ) | < 0.3% (14 MeV) |
| | < 0.5% (6 MeV) |
| Nor. RMS emittance (effective part ) | < 16 mm mrad |
| Beam length (effective part) | < 10 ps |
| Micro-pulse effective charge | > 200 pC |

*Corresponding author (email: lshang@ ustc.edu.cn)
†Work supported by NSFC No. 10875116

## 2  The 14 MeV injector

### 2.1  EC-ITC RF gun

The ITC (Independently-Tunable-Cells) RF gun can compress the bunch well and get expected beam quality by setting the appropriate feeding power and launch phases separately, instead of using α-magnet or complicated laser drive system. In addition, the EC (External-Cathode) structure we use can increase the captured beam current and decrease the energy spread than the common structure with cathode inside the cavity. And the negative effect of back bombardment to the cathode is almost eliminated [5,6].

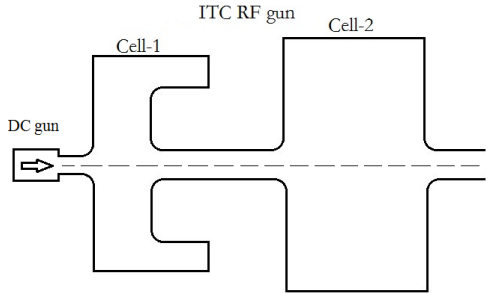

Figure 1: Layout of the external injecting ITC RF gun.

### 2.1.1  15 KeV gridded DC gun

15 KeV gridded DC gun was designed to provide pulsed beam. In the preliminary design, a special geometry with intermediate electrode was used, which is called double-anode structure, to provide 4.2A beam current [7]. In order to enhance the electric field strength on the emitting surface and the transverse focusing force, an intermediate electrode was added to accelerate the electrons to higher energy, and then the electrons will be decelerated to ~15 KeV between intermediate electrode and the anode.

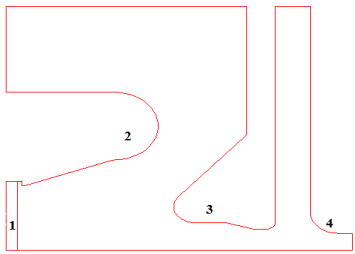

Figure 2: Model of 15 KeV DC gun. From 1 to 4, the electrodes are emitter, focusing electrode, intermediate electrode and anode.

After times of optimizations, the lower–current beam injected into the ITC RF gun can also satisfy the requirements of beam quality at the injector exit. This structure makes it possible to provide laminar beam at different currents in certain range though adjust the high voltage on the intermediate electrode. In this paper, the beam current we need is 2.45A. The OPERA [8] model of the gun is shown in Fig. 3. The maximum electric field in the gun is 8.9 MV/m, which is lower than the breakdown threshold.

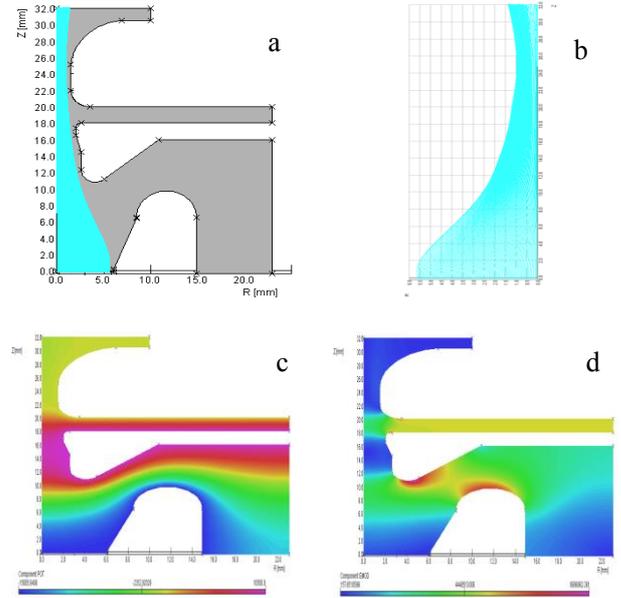

Figure 3: a. OPERA model with trajectory b. Beam trajectory c. Potential distribution d. Electric field strength distribution

The simulation results were verified with CST Particle Studio [9]. The trajectory of the electrons in Fig. 4 is nearly the same with that from OPERA.

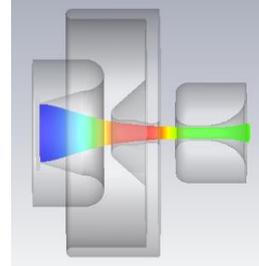

Figure 4: Beam trajectory in the 3D model from CST

### 2.1.2  ITC RF gun

The beam from 15 KeV DC gun is captured and bunched in the first cell of ITC, and then the beam bunches are compressed and accelerated to ~2.3 MeV in the second cell. The two cells are power fed independently and achieve resonance at 2856 MHz, which are modeled with Superfish code [10].

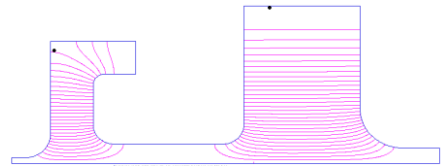

Figure 5: Superfish model of the ITC RF gun

Dynamics computation from the exit of DC gun to the entrance of beam transport line was executed with ASTRA code [11].The initial distribution was generated according to the simulation results from OPERA and CST, as shown in Fig. 6.

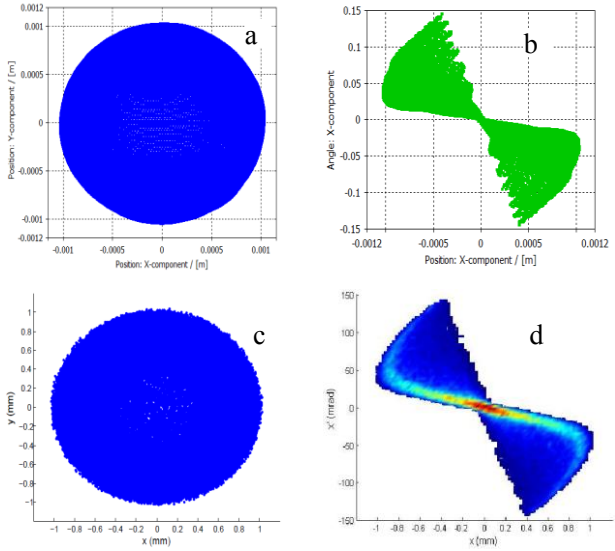

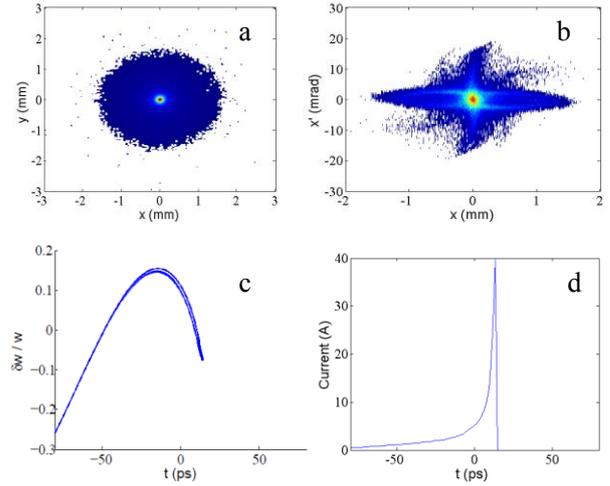

Figure 6: a. Transverse size of the beam at the exit of DC gun from CST.  b. Transverse phase space of the beam at the exit of DC gun from CST.  c. Transverse size of the beam for dynamic simulation.  d. Transverse phase space of the beam for dynamic simulation.

Since the ITC RF gun consists of two independently tunable cells, the parameters of the two cells, including peak electric fields and the phases needs to be optimized to match. The peak field electric fields on the axis are 42 MV/m in the first cell and 90 MV/m in the second cell.

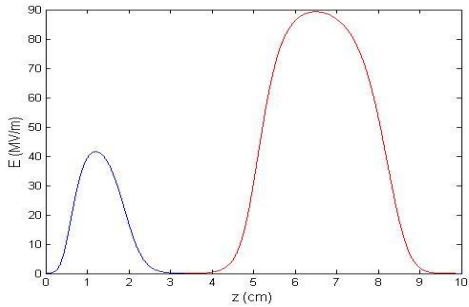

Figure 7: Magnitude of electric field on axis

Due to the advantages of ITC RF gun, we can provide expected bunch at different launching phases by matching the setting phases of cell one and cell two. Optimizations have been made to get the best output. The particle distribution at the entrance of travelling-wave structure is shown in Fig. 8.

Figure 8: Particle distribution at the TWS entrance
a. Transverse size         b. Transverse phase space
c. Longitudinal phase space.   d.  Longitudinal distribution

### 2.2   Compensating coils

In order to compensate the emittance growth, a short lens and a solenoid were designed [11].The center of the short lens is 192 mm downstream the cathode. The magnetic field is optimized cell by cell to get the best beam quality.

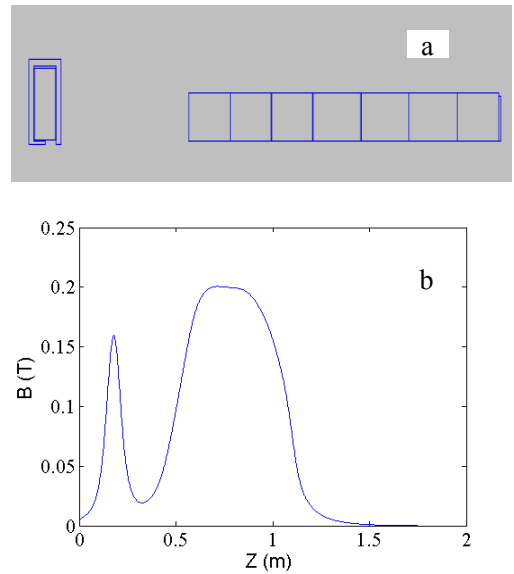

Figure 9: a：Model of short magnetic lens and solenoid

b: Optimized magnetic field on the axis (14 MeV)

### 2.3   The Travelling-wave structure (TWS)

The constant gradient travelling-wave structure consists of one input coupler, 19 accelerating cells with $2\pi/3$ mode and  4  collinear  absorbing  loads  which  coating wave-absorbing materials on cavity inner surfaces.

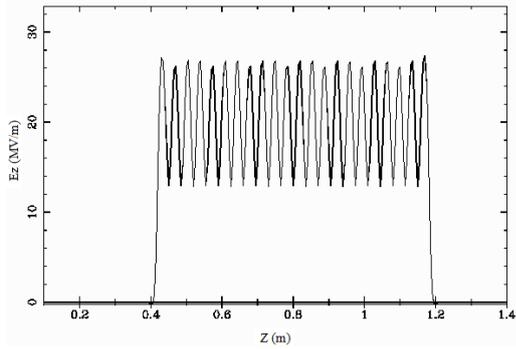

Figure 10: Electric field on the axis in the TWS

In this simulation, the travelling-wave structure is represented by an input coupler, 21 travelling-wave cells and an output coupler. The maximum field on the field is 27.65 MV/m, and beam loading effect is also considered with ASTRA code. Particle distributions at the 90mm downstream of injector are shown in Fig.11.

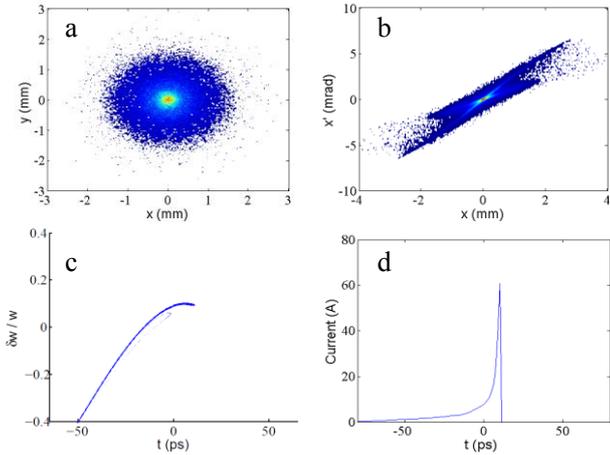

Figure 11: Particle distribution at the injector exit
a. Transverse size  b. Transverse phase space
c. Longitudinal phase space.  d. Longitudinal distribution

### 2.4 Analysis of simulation results

The performance of output is characterized with the parameters of effective part (the head) of the bunch, of which RMS energy spread should be less than 0.3% within 10ps. The effective part determines the quality of the THz radiation in the undulator, of which the charge should be more than 200 pC. Moreover, the emittance of effective part should be less than 16 mm mrad.

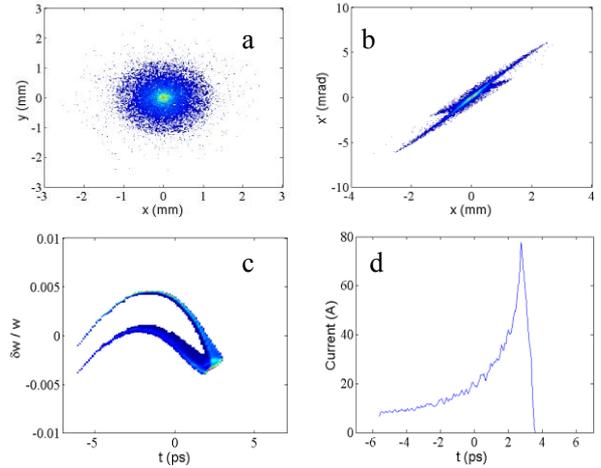

Figure 12: Particle distribution of the effective part
a. Transverse size  b. Transverse phase space
c. Longitudinal phase space.  d. Longitudinal distribution

As shown in Tab.2, the emittance, energy spread and charge of effective part can satisfy the strict requirements of the THz radiation. The emittance is 14.84 mm mrad, and the energy spread is 0.26%.

Table 2  Properties of output beam from the injector
(14 MeV)

| | |
|---|---|
| Beam energy (effective part) | 13.91 MeV |
| Beam current (micro pulse) | 60 A |
| Micro-pulse length (FWHM) | 3.8 ps |
| Energy spread (effective part) | 0.26 % |
| Nor. RMS emittance (effective part) | 14.84 mm mrad |
| Beam length (effective part) | 9 ps |
| Micro-pulse effective charge | 202 pC |

### 3  The 6 MeV operating mode

It is expected that the injector can provide the beam bunch with the energy in the range of 6 MeV to 14 MeV. Thus, the injector at 6 MeV mode has also been simulated and optimized.

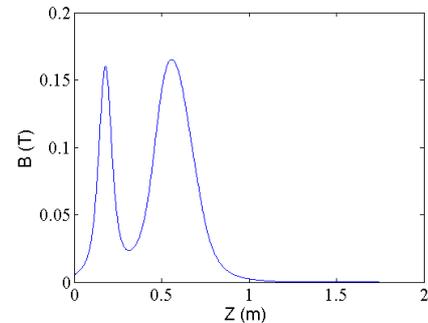

Figure 13: Optimized magnetic field on the axis (6 MeV)

At this mode, the properties of beam from EC-ITC RF gun should be the same with that in 14 MeV mode. The magnetic field has been adjusted to fit 6 MeV case, which is depicted is Fig. 13. In order to accelerate the bunch to 6 MeV, the maximum field in the travelling-wave structure should be 12.65 MV/m. The beam distribution of effective part at the injector exit is shown in Fig. 14.

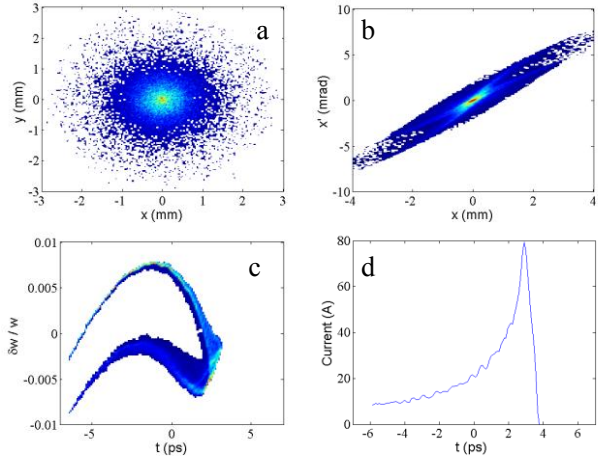

Figure 14: Particle distribution of the effective part (6 MeV)
a. Transverse size  b. Transverse phase space
c. Longitudinal phase space.  d. Longitudinal distribution

The parameters in Tab. 2 show that the output beam of the injector achieves the anticipated target. When operating at this mode, the emittance is 14.2 mm mrad and the energy spread is 0.449%.

Table 3   Properties of output beam from the injector (6 MeV)

| | |
|---|---|
| Beam energy (effective part) | 5.98 MeV |
| Beam current (micro pulse) | 50 A |
| Micro-pulse length (FWHM) | 5.71 ps |
| Energy spread (effective part) | 0.449 % |
| Nor. RMS emittance (effective part) | 14.2 mm mrad |
| Beam length (effective part) | 9 ps |
| Micro-pulse effective charge | 201 pC |

## 4  Multi-bunch simulations

Muti-bunch simulations have been executed to study the interaction of each bunch. Taking the 14 MeV as an example, we can see that tail of the ahead bunch becomes a part of the head part next bunch in Fig. 15. This phenomenon will cause the emittance growth and energy spread growth of the next bunch. Nevertheless, the analysis results show that the effective part is still high-quality, of which the emittance is less than 15 mm mrad with more than 200pC charge. This phenomenon also works on 6 MeV mode.

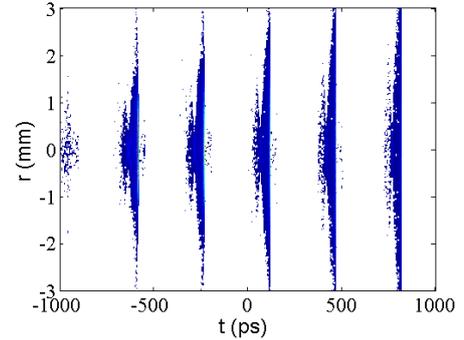

Figure 15: Longitudinal distribution of multi-bunch case

## 5  Conclusion

This paper has described the injector for a compact THz radiation source. Start-to-end simulations have been done to check the design target. Laminar beams at different current values in certain range can be provided by the 15 KeV gridded DC gun. The EC-ITC RF gun can generate desired output through various combinations of phases in cell one and cell two. The output beam from the travelling-wave structure can satisfy the requirements of THz source. The effective charge is more than 200pC, and the normalized RMS emittance is less than 15 mm mrad. The energy spread is less than 0.3% at 14 MeV, which is less than 0.5% at 6 MeV.